\documentclass[preprint,showpacs,preprintnumbers,amsmath,amssymb]{revtex4}

\usepackage{epsfig}
\usepackage{graphicx}
\usepackage{dcolumn}
\usepackage{bm}

\newcommand{\ord}{{\cal O}}
\def\beq{\begin{equation}}
\def\eeq#1{\label{#1}\end{equation}}
\def\eeqn{\end{equation}}
\newcommand\iden{\leavevmode\hbox{\small1\normalsize\kern-.33em1}}

\let\jnfont=\rm
\def\NPB#1,{{\jnfont Nucl.\ Phys.\ B }{\bf #1},}
\def\PLB#1,{{\jnfont Phys.\ Lett.\ B }{\bf #1},}
\def\EPJC#1,{{\jnfont Eur.\ Phys.\ Jour.\ C }{\bf #1},}
\def\PRD#1,{{\jnfont Phys.\ Rev.\ D }{\bf #1},}
\def\PRL#1,{{\jnfont Phys.\ Rev.\ Lett.\ }{\bf #1},}
\def\MPLA#1,{{\jnfont Mod.\ Phys.\ Lett.\ A }{\bf #1},}
\def\JPG#1,{{\jnfont J.\ Phys.\ G }{\bf #1},}
\def\CTP#1,{{\jnfont Commun.\ Theor.\ Phys.\ }{\bf #1},}
\def\JHEP#1,{{\jnfont JHEP \ }{\bf #1},}
\def\NPPS#1,{{\jnfont Nucl.\ Phys.\ Proc.\ Suppl.\ }{\bf #1},}
\def\CPC#1,{{\jnfont Computl.\ Phys.\ Commun.\ }{\bf #1},}

\begin{document}

\preprint{\parbox{1.2in}{\noindent }}

\title{\ \\[10mm]  Higgs boson productions at LHC as a probe of
                   different littlest Higgs models with T-parity}

\author{\ \\[2mm] Lei Wang, Jin Min Yang \\ ~}

\affiliation{Institute of Theoretical Physics, Academia Sinica,
             Beijing 100080, China \vspace*{1.5cm}}

\begin{abstract}
Higgs boson productions at the LHC will serve as a sensitive probe
of various little Higgs models. In this work we comparatively study
two littlest Higgs models with different T-parity
constructions through examining their effects in three production
processes of the Higgs boson at the LHC, namely the productions of a
single Higgs, a Higgs-pair, as well as a Higgs boson associated with
a pair of top and anti-top quarks. The two models are characterized
by predicting a top partner canceling the Higgs mass quadratic
divergence contributed by the top quark with even and odd T-parity,
respectively. We find that both models can alter the SM cross sections
 sizably and their corrections also differ significantly.
Therefore, the Higgs boson productions at the LHC may shed some light 
on these two models or even distinguish them. 
\vspace*{1cm}
\end{abstract}


\maketitle

\section{Introduction}
Little Higgs theory \cite{ref1} has been proposed as an interesting
solution to the hierarchy problem. In such a theory the Higgs boson
is a pseudo-Goldstone boson and its mass is protected by an
approximate global symmetry and free from one-loop quadratic
sensitivity to the cutoff scale. The littlest Higgs model
\cite{ref2} economically implements the idea of the little Higgs
theory.  Due to the tree-level mixing of heavy and light mass
eigenstates, the electroweak precision tests can give strong
constraints on the model \cite{ref3}, which would require raising
the mass scale of the new particles to be much higher than TeV and
thus reintroduce the fine-tuning in the Higgs potential \cite{ref4}.
To tackle this problem, a discrete symmetry called T-parity has been
proposed \cite{ref5}, which forbids those tree-level contributions.
In the pioneer version of such model  (hereafter called model-I)
\cite{ref5}, the T-parity is simply implemented by adding the
T-parity images for the original top quark interaction to make the
Lagrangian T-invariant. A characteristic prediction of this model is
a T-even top partner which cancels the Higgs mass quadratic
divergence contributed by the top quark. Since the heavy top partner
is T-even, it can be singly produced at the LHC, which is a crucial
phenomenology of this model.

An alternative implementation of T-parity has recently been proposed
(hereafter called model-II)  \cite{ref6},  where all new particles
including the heavy top partner responsible for canceling the SM
one-loop quadratic divergence are odd under T-parity. An obvious
virtue of this model is that the spectrum of the third-generation
quark sector is simplified \cite{ref6}. Many studies of the collider
phenomenology for model-I have been done \cite{model-I}. However,
the phenomenology of model-II is quite different from model-I
\cite{ref6}, especially for the heavy top partner, which is T-odd
and cannot be singly produced at the LHC.

To probe these models at the LHC, the Higgs boson production
processes are ideal places. Firstly, these littlest Higgs models
mainly alter the Higgs sector of the SM and thus the Higgs
properties may deviate from the SM Higgs boson. Secondly, the Higgs
boson is the most important target of the LHC experiment \cite{ref8}
and its various production channels will be explored at the LHC. In
this work we choose three typical production processes of the Higgs
boson at the LHC as a probe of these littlest Higgs models. The
first one is the production of a single Higgs boson via gluon-gluon
fusion, which is the dominant production mechanism at the LHC
\cite{ref9}. The second one is the Higgs-pair production, which is
rare but very important since it provides a way to probe the Higgs
boson self-coupling \cite{ref10}. The third process is the
production of a Higgs boson in association with a pair of top
quarks, which plays an important role in testing the Yukawa coupling
\cite{ref11}.  These processes have been studied in model-I
\cite{ref12,ref13,ref14,ref15}, but not yet in model-II. In this
work, we comparatively study the effects of both models in these
three Higgs production processes.

This work is organized as follows. In Sec. II we recapitulate the
fermion and top quark Yukawa sector of the models. Since model-I
has been elucidated in detail in the literature, we focus on   
model-II.  In Sec. III, we
study the effects of these models in the productions of a single
Higgs, a Higgs-pair and a Higgs boson in association with a top
quark-pair at the LHC. Finally, we give our conclusion in Sec. IV.

\section{The littlest Higgs models with T-parity}

The original littlest Higgs model \cite{ref2} is based on a
non-linear sigma model describing the spontaneous breaking of a
global $SU(5)$ down to a global $SO(5)$ at an energy
scale $f\sim\ord(TeV)$. The vacuum expectation value (VEV) of an
$SU(5)$ symmetric tensor $\Sigma$ is proportional to \beq \Sigma_0
\,=\, \left(\begin{array}{ccc}
0& 0& \iden\\
0& 1& 0\\
\iden& 0& 0\\
\end{array}\right),
\eeq{sigma0} where $\iden$ represents a unit $2\times 2$ matrix. The
low energy dynamics of non-linear sigma is described in terms of the
field \beq \Sigma(x) \,=\, e^{i \Pi/f} \Sigma_0 e^{i \Pi^T/f} \,=\,
e^{2i \Pi/f} \Sigma_0
 \eeq{sigma_def}
 with
\beq \Pi(x)=\sum_{a=1}^{14} \pi^a(x) X^a,
 \eeq{pionmatrix}
where $\pi^a(x)$ are the Goldstone particles corresponding to 14
broken generators $X^a$ for the $SU(5)\to SO(5)$ breaking.

In the pioneer version of littlest Higgs model with T-parity
(model-I), the T-parity in the top quark sector is implemented by
simply adding the T-parity images of the original interaction to
make the Lagrangian T-invariant. Thus, the heavy top partner which
cancels the Higgs mass quadratic divergence contributed by the top
quark is T-even. There are detailed descriptions of this model in
the literature \cite{ref5} and we do not discuss it in detail here.
In the following, we recapitulate an alternative version of T-parity
construction (model-II) \cite{ref6}.

In model-II for each generation of fermion (quark and lepton), we introduce two
doublets $q_1$ and $q_2$, which are embedded into the incomplete representations of $SU(5)$
multiplets $Q_1$ and $Q_2$, and a right-handed $SO(5)$ multiplet $\Psi_R$ which transforms
nonlinearly under the full $SU(5)$.
The field content can be expressed as
\begin{equation}
\begin{array}{ccc}
Q_1=\left(\begin{array}{c} q_1 \\ 0 \\ 0 \end{array}\right)\,,
& Q_2=\left(\begin{array}{c} 0 \\ 0 \\ q_2
\end{array}\right) \,,&
\Psi_R=\left(\begin{array}{c} \psi_R \\ \chi_R \\ \tilde{\psi}_R
\end{array}\right),
\end{array}
\end{equation}
where $q_A=\left( -i d_{LA}, i u_{LA} \right)^{\rm T}$ with $A=1,
2$, and $\psi_R=\left( -i d'_{R}, i u'_{R} \right)^{\rm T}$. The
first component of the $\tilde{\psi}_R$ is  irrelevant to our study
(as shown later), and the second component of the $\tilde{\psi}_R$
is $ iq_R$. The mirror fermions can be given $\ord(f)$ masses via a
mass term
 \cite{ref6},
\begin{equation}\label{heavyyuk}
{\cal L}_{\kappa}=-\kappa_{ij}  f (\bar{Q}^i_1 \xi -\bar{Q}^i_2
\Sigma_0 \Omega \xi^\dagger)\Psi^j_R +h.c.,
\end{equation}
where $\xi=e^{i\Pi/f}$, $\Omega \equiv {\rm diag}(1,1,-1,1,1)$, and
$i$, $j=1, 2, 3$ are the generation indices. For simplicity, we
assume the flavor diagonal and universal $\kappa$ in the study.

They transform under the $SU(5)$ as \beq Q_1 \rightarrow V Q_1\,,
\hspace{.2in} Q_2 \rightarrow V^* Q_2\,, \hspace{.2in}\Psi_R
\rightarrow U\Psi_R, \hspace{.2in} \xi \rightarrow V\xi U^\dagger,
\hspace{.2in} \Sigma \rightarrow V\Sigma V^{\rm T}, \eeq where $V$
is an $SU(5)$ rotation matrix, and $U$ is the unbroken $SO(5)$
rotation and is a non-linear representation of the $SU(5)$. Under
T-parity, the transformation laws are defined as
\beq Q_1
\leftrightarrow \Sigma_0 Q_2,  \hspace{.2in} \Psi_R \rightarrow
-\Omega\Psi_R,  \hspace{.2in} \xi \rightarrow \Omega \xi^\dagger
\Omega.
\end{equation}
 Thus $q_1\leftrightarrow q_2$,  and $\Sigma \rightarrow
\tilde{\Sigma}=\Sigma_0 \Omega \Sigma^\dagger \Omega \Sigma_0$ under
T-parity.  Following the above transformation, the Lagrangian is
T-invariant.

The Lagrangian in Eq. (5) contains the new Higgs boson interactions
and the mass terms. For the first and second generations,
\begin{eqnarray}
{\cal L}_{\kappa} &\simeq&-\sqrt{2} \kappa f \left[ \bar{d}_{L_-}
d'_{R}+\frac{1+c_\xi}{2} \bar{u}_{L_-} u'_R
-\frac{1-c_\xi}{2}\bar{u}_{L_-} q_R +\frac{s_\xi}{\sqrt{2}}
\bar{u}_{L_+} \chi_R
 \right]+{\rm h.c.},
\label{Kappa_int}
 \end{eqnarray}
where we ignored the generation indices, and
 $c_\xi (\equiv \cos\frac{v+h}{\sqrt{2}f})$ and
$s_\xi (\equiv \sin\frac{v+h}{\sqrt{2}f})$ come from the non-linear
sigma model field $\xi$, with  $h$ and $v$ being the neutral Higgs
boson field and its VEV, respectively. The fermion $u_{L_{-}} =
(u_{L_1}- u_{L_2})/\sqrt{2}$ is T-odd, which together with $u'_R$
gets a mass, and $u_{L_{+}} = (u_{L_1}+ u_{L_2})/\sqrt{2}$ is T-even
and massless. The same definition also applies to the down-type
quarks. The fields $q_R$ and $\chi_R$ can be given large Dirac
masses by introducing additional fields, as described in detail in
\cite{ref5}.  We will simply assume that their masses  are
$5f$. From the above Eq. (8), we can see the the first component of
the doublet $\tilde{\psi}_R$ doesn't appear  and the T-odd down-type
quarks have no tree-level coupling with Higgs boson. 

For the top quark sector, in order to cancel the quadratic divergence 
of the Higgs mass induced by top quark, it requires the introduction of the
additional singlets as follows: $Q_1=(q_1,U_{L1},0_2)^{\rm T}$ and
$Q_2=(0_2,U_{L2},q_2)^{\rm T}$. From Eq. (5) we can get the Higgs
boson interactions and the mass terms for the third generation
fermions
\begin{eqnarray}
{\cal L}_{\kappa} &\simeq&-\sqrt{2} \kappa f [ \bar{d}_{L_-}
d'_{R}+\frac{1+c_\xi}{2} \bar{u}_{L_-} u'_R
-\frac{1-c_\xi}{2}\bar{u}_{L_-}q_R
-\frac{s_\xi}{\sqrt{2}}\bar{U}_{L_-}q_R
-\frac{s_\xi}{\sqrt{2}}\bar{U}_{L_-}u'_R \nonumber \\
&&+\frac{s_\xi}{\sqrt{2}}\bar{u}_{L_+} \chi_R +
c_\xi\bar{U}_{L_+}\chi_R ]+{\rm h.c.},
 \end{eqnarray}
where the T-parity eigenstates are defined as $U_{L_{+}} = (U_{L_1}+
U_{L_2})/\sqrt{2}$ (T-even),  and $U_{L_{-}} = (U_{L_1}-
U_{L_2})/\sqrt{2}$ (T-odd). The $U_{L_+}$  together with $\chi_R$
gets a Dirac mass.

Introducing additional singlets $U^c_1 \leftrightarrow U^c_2$
under T-parity, the top quark Yukawa coupling can be written as
\cite{ref6}
\begin{equation}
{\cal L}_t= -\frac{\lambda}{2}f\epsilon_{ijk} \epsilon_{xy}
\left[(Q_1)_i \Sigma_{jx} \Sigma_{ky}U^c_1+ (\Sigma_0Q_2)_i
\tilde{\Sigma}_{jx} \tilde{\Sigma}_{ky}U^c_2 \right] +{\rm h.c.},
\end{equation}
where the indices $i$, $j$, $k$ run from 1 to 3 whereas $x$, $y$=4,
5. The Eq. (10) will introduce mixing between the light T-even and
the heavy T-even fermions, which can be removed by the additional
interactions \cite{ref6},
\begin{equation}
{\cal L}'_t= -\frac{\lambda'}{2}f\epsilon_{lmn} \epsilon_{rs}
\left[(\Omega Q_2)_l \Sigma'_{mr} \Sigma'_{ns}U^c_1+(\Omega\Sigma_0
Q_1)_l \tilde{\Sigma'}_{mr} \tilde{\Sigma'}_{ns}U^c_2 \right]+{\rm
h.c.},
\end{equation}
where the indices $l$, $m$, $n$ run from 3 to 5 whereas $r$, $s$=1,
2. $\Sigma'=\Omega \Sigma^\dagger \Omega$, and
$\Sigma'\rightarrow\tilde{\Sigma}'=\Sigma_0\Sigma\Sigma_0$ under
T-parity. Adding  ${\cal L'}_t$ to ${\cal L}_t$, and taking
$\lambda'=\lambda$, we can get the following simple expression of
top quark Yukawa sector,
\begin{eqnarray}
{\cal L}_t +{\cal L}'_t&\simeq& -\lambda f \left(\sqrt{2}s_\Sigma
u_{L_+}U^c_+ +(1+c_\Sigma) U_{L_-} U^c_- \right)+{\rm h.c.},
\end{eqnarray}
where $c_\Sigma(\equiv\cos\frac{\sqrt{2}(v+h)}{f})$ and
$s_\Sigma(\equiv \sin\frac{\sqrt{2}(v+h)}{f})$ are originated from
the non-linear sigma model field $\Sigma$, and
$U^c_+=(U^c_1+U^c_2)/\sqrt{2}$, $U^c_-=(U^c_1-U^c_2)/\sqrt{2}$.

The Yukawa couplings of up-type quarks for the first and second generations are
given by the similar Lagrangian for the top quark,  but without introducing extra
singlet fields,
\begin{equation}
{\cal L}_u= -\frac{\lambda_u}{2}f\epsilon_{ijk} \epsilon_{xy}
\left[(Q_1)_i \Sigma_{jx} \Sigma_{ky}u^c+ (\Sigma_0Q_2)_i
\tilde{\Sigma}_{jx} \tilde{\Sigma}_{ky}u^c \right] +{\rm h.c.},
\end{equation}
where $u^c\rightarrow u^c$ under T-parity. The Eq. (13) contains the
following Higgs boson interactions as well as the mass term for
up-type quarks of the first and second generations,
\begin{equation}
{\cal L}_u \simeq -\lambda_u f s_\Sigma u_{L_+}u^c+{\rm h.c.}.
\end{equation}
After diagonalizing the mass matrix Eq. (8, 9, 12, 14), we can get
the mass eigenstates of new fermions. For each  SM fermion doublet,
there are $d_-$, $u_-$ , $q$ (T-odd), and $\chi$ (T-even).  Besides,
the top quark has a T-odd  partner $T_-$ which cancels the one loop
quadratic divergence of Higgs mass induced by top quark.

\section{The effects in Higgs boson productions at the LHC}

The relevant Feynman rules in our calculations can be obtained after
diagonalizing the mass matrix in Eqs. (8, 9, 12, 14), which is
performed numerically in our analyses. The Higgs mass is assumed to
be 150 GeV and other SM parameters involved are taken as $m_t=172.7$
GeV, $m_Z=91.187$ GeV \cite{ref16}, and we use the two-loop running
coupling constant $\alpha_{s}(Q)$ with $\alpha_{s}(m_{Z})=0.118$.
For the parton distributions we use CTEQ6L \cite{ref17} with the
renormalization scale $\mu_R$ and the factorization scale $\mu_F$
chosen to be $\mu_R=\mu_F=m_h$ for single Higgs production and
$\mu_R=\mu_F=2m_h$ for Higgs-pair production. To simplify the top
quark Yukawa sector, we have taken Yukawa coupling constants
$\lambda'=\lambda$. Therefore, the new free parameters involved are
the breaking scale $f$ and $\kappa$. Our calculations show that the
results are not very sensitive to $\kappa$ in the allowed parameter
space. Thus, we take $\kappa=3.0$, and retain only $f$ as free
parameter. The \cite{ref18} shows that the scale $f$ in model-I may
be below 1 TeV, and the constraint in model-II is expected to be
even weaker \cite{ref6}.

Our calculations will deal with loop diagrams.
The calculations of such loop diagrams are straightforward.
Each loop diagram is composed of some scalar loop
functions \cite{ref20} which are calculated by using LOOPTOOLS
\cite{ref21}. Since the explicit expressions of these form factors
are lengthy, we will not present them in the paper.
\begin{figure}[htb]
\epsfig{file=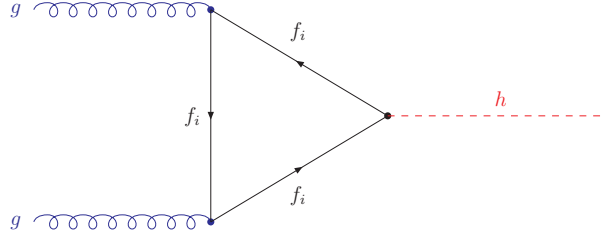,width=8cm}
\vspace*{-0.5cm}
 \caption{\small The main parton-level Feynman diagrams for single Higgs boson
production via gluon-gluon fusion in model-II. Here, $f_i$=t,
$\chi$, $T_-$, $u_-$, $q$  for the third generation,
 and $f_i$=$\chi$, $u_-$, $q$ for the first two generations.}
\label{fig1}
 \end{figure}
\begin{figure}[htb]
\epsfig{file=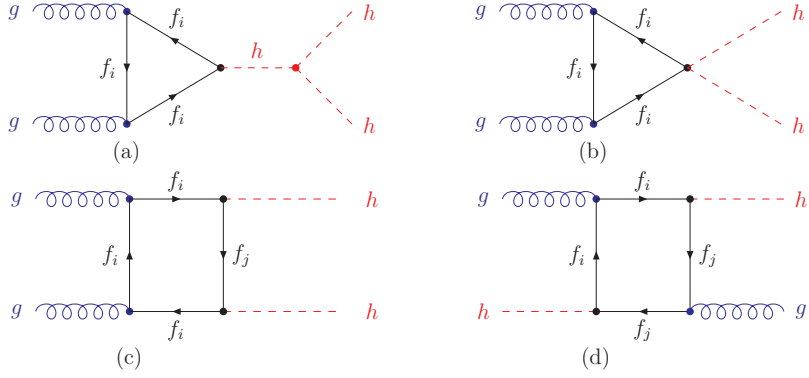,width=11cm}
\vspace*{-0.5cm}
\caption{\small The parton-level Feynman diagrams for Higgs-pair production
  via gluon-gluon fusion in model-II.
  Here, $f_i$ can be a T-even quark ($i=1, 2$ with $f_1=u$ and $f_2=\chi$
  for three generations) or a T-odd quark (for the third generation,
  $i=1, 2,  3$ with $f_1=T_-$, $f_2=u_-$ and $f_3=q$;
  while for the first and second generations, $i=1, 2$ with  $f_1=u_-$ and $f_2=q$).
  The other diagrams obtained by exchanging the two gluons or exchanging the two Higgs
  bosons are not shown here.}
\label{fig2}
 \end{figure}
\begin{figure}[htb]
\epsfig{file=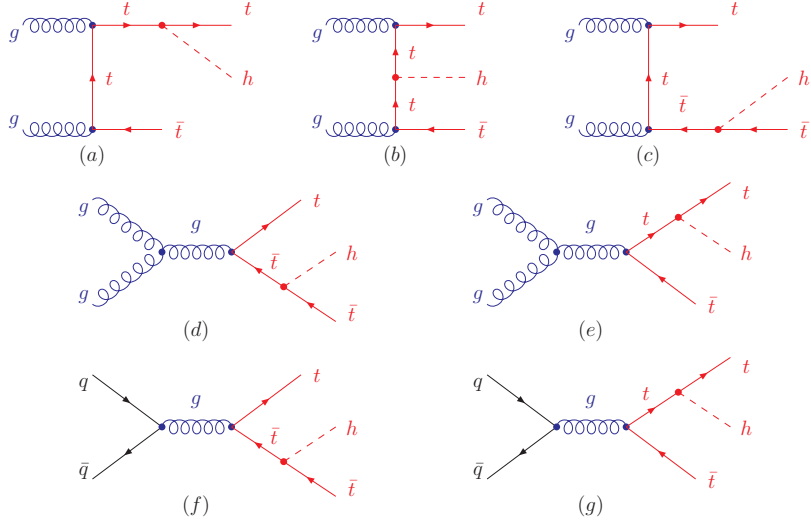,width=11cm}
\vspace*{-0.5cm}
\caption{The patron-level Feynman diagrams for $ht\bar{t}$ production at the LHC in model-II.
 The u-channel diagrams by exchanging the two gluons in (a)-(c) are not shown here.}
\label{fig3}
\end{figure}

At the LHC the single Higgs production via gluon-gluon fusion
is dominated by top quark loop in the SM. In model-II, the
Feynman diagrams of the process are shown in Fig. \ref{fig1}. Due to the
modified $ht\bar{t}$ coupling, the top quark loop may give some
corrections to SM prediction. In addition to the top quark loop, the
loops of new T-even and T-odd quarks also come into play.
In model-I the corrections are also mainly from these two aspects.
The relevant Feynman diagrams and rules are described in detail in \cite{ref12}.

The Higgs-pair production at the LHC can proceed through gluon-gluon
fusion and $b\bar{b}$ annihilation at parton level, with the former
being the dominant one \cite{ref19}. The main Feynman diagrams of
the process $gg\to hh$ in model-II are shown in Fig. \ref{fig2}. In
the SM the dominant contributions are from the diagrams in Fig.
\ref{fig2} (a, c, d) with top-quark running in the loops. In
model-II the top quark can give the new correction through the
tree-level $hht\bar{t}$ coupling and the modified $ht\bar{t}$
coupling. Also, since the Higgs boson interacts with the T-even and
T-odd quarks introduced, we have additional diagrams with these new
quarks running in the loops. In model-I the diagrams of the
corrections are similar, which are presented in \cite{ref14}. In our
calculation, we ignore the contributions of the light quark loops.

The production of $h t\bar t$ at the LHC can proceed through $gg$
fusion and $q\bar{q}$ annihilation at parton level. In both model-II
and model-I, the tree-level Feynman diagrams are the same as in the
SM, which are shown in Fig. \ref{fig3}. The corrections are mainly
from the modified $h t\bar t$ coupling.

\begin{figure}[htb]
\epsfig{file=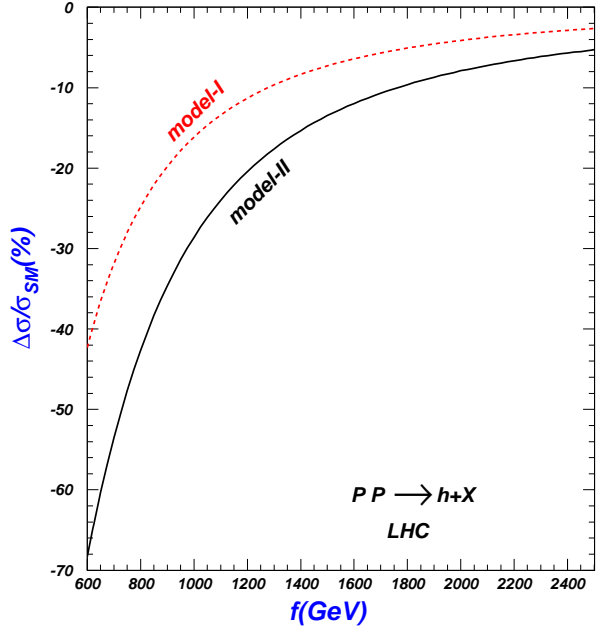,width=8cm} \vspace*{-0.5cm} \caption{The
corrections to the SM single Higgs boson production rate versus the
parameter $f$  at the LHC.}

\label{fig4}
\end{figure}
\begin{figure}[htb]
\epsfig{file=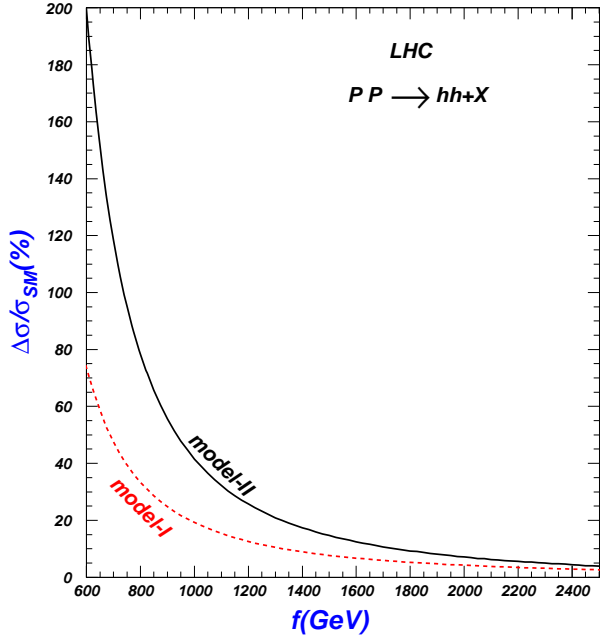,width=8cm}
\vspace*{-0.5cm}
\caption{\small The corrections to the SM Higgs-pair
   production rate versus the parameter $f$  at the LHC.}
\label{fig5}
\end{figure}
\begin{figure}[htb]
\epsfig{file=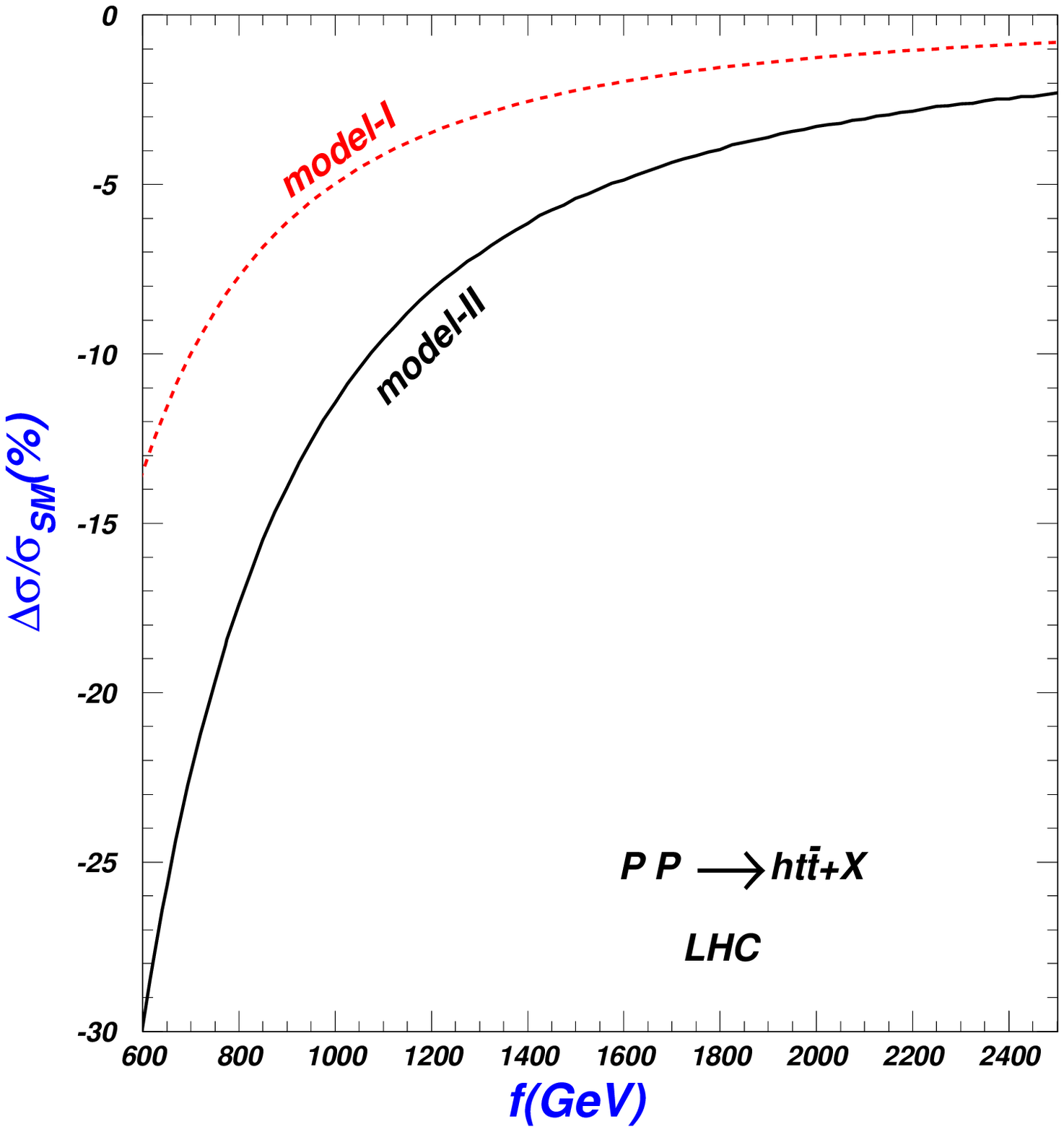,width=8cm}
\vspace*{-0.5cm}
\caption{The corrections to the SM $ht\bar{t}$ production rate
versus the parameter $f$  at the LHC.}
\label{fig6}
\end{figure}

In Figs. \ref{fig4}-\ref{fig6} we plot the corrections to the SM
predictions of the production rates versus the parameter $f$. In
model-I we take $r=1.0$ and our results agree with
\cite{ref12,ref14,ref15}.

Figs. \ref{fig4}-\ref{fig6} show that the contributions of these two
models can significantly alter the SM cross sections in the allowed
parameter space. The corrections are sensitive to the scale $f$ and
the magnitude becomes more sizable for lower values of $f$.
Furthermore, we see the corrections in model-II are much more
sizable than model-I. For example, for model-II (model-I) the
correction is  -43\% (-25\%) with $f=800$ GeV  in Fig. \ref{fig4},
41\% (19\%) with $f$=1TeV in Fig. \ref{fig5},  and -30\%(-13.5\%)
with $f=600$ GeV in Fig. \ref{fig6}.

\section{Conclusion}
In this work we comparatively studied
two typical littlest Higgs models with different T-parity constructions
through examining their effects in three production processes
of the Higgs boson at the LHC, namely the productions of a single
Higgs, a Higgs-pair, as well as a Higgs boson associated with a
pair of top and anti-top quarks.
We found that both models can alter the SM cross sections
sizably and their corrections also differ significantly.
Therefore, the Higgs boson productions at the LHC may
shed some light on these two models or even distinguish them.

\section*{Acknowledgement}
This work was supported by National Natural Science
Foundation of China (NNSFC) under No. 10475107, No. 10725526 
and No. 10635030.

\end{document}